\documentclass[11pt]{article}
\usepackage{moriond,epsfig}
\usepackage{amsmath}
\usepackage{amsfonts}
\usepackage{amssymb}

\def\Journal#1#2#3#4{{#1} {\bf #2}, #3 (#4)}

\def\PRL{\em Phys. Rev. Lett.}

\def\be{\begin{equation}}
\def\ee{\end{equation}}
\def\bea{\begin{eqnarray}}
\def\eea{\end{eqnarray}}

\begin{document}
\vspace*{4cm}
\title{HEAVY FLAVOUR PHYSICS AT CMS AND ATLAS}

\author{ L. WILKE\\
	on behalf of the CMS and ATLAS Collaborations }

\address{Physik-Institut, Universit\"at Z\"urich, 
	Switzerland}

\maketitle\abstracts{
	Prospects for heavy flavour studies with the CMS and ATLAS detectors are presented. Many studies are aimed for early LHC data, taking advantage of the large $b$ production cross-section. Rare decay studies as the $B_s \rightarrow \mu^+\mu^-$ decay have also been performed.  
}
\section{Introduction}
CMS~\cite{:2008zzk} and ATLAS~\cite{:2008zzm} are multipurpose detectors operating at the LHC at CERN. Their excellent tracking and muon systems up to high pseudorapidity makes them well suited for heavy flavour studies. Since a high number of charm and beauty quarks will be produced at the LHC, analyses will already be possible with integrated luminosities of 10 pb$^{-1}$ . If not mentioned otherwise, the studies presented assume a centre-of-mass energy of 14 TeV and are made with full detector simulations.

The outline is as follows. In Section \ref{sec:quarkonia} the strategies for quarkonia studies including cross-section and polarisation measurements are discussed. Section \ref{sec:bprod} covers $b$-quark production whereas in Section \ref{sec:bdecay} the measurement prospects for different decays of $b$-mesons are discussed.

\section{Quarkonia studies}
\label{sec:quarkonia}
Several models exist for the production mechanism of quarkonium \cite{Brambilla:2004wf,Lansberg:2006dh,Kramer:2001hh}. While the color-octet mechanism describes well the inclusive quarkonium cross-section at the Tevatron, it does not describe the polarisation \cite{cdfoniapol}. Other models have been proposed but it is not clear yet as to which describes the data best. Both CMS and ATLAS have prepared analyses to probe the region of large transverse momenta with high statistics accessible only at the LHC to improve on understanding of the production mechanism. Furthermore, quarkonia studies are vital for detector alignment and calibration.

\subsection{$J/\psi$-cross-section measurement}
There are three main sources of $J/\psi$ production, directly produced $J/\psi$, prompt $J/\psi$ produced indirectly (e.g. from $\chi_c$ decays) and non-prompt $J/\psi$ from the decay of $b$ hadrons. CMS studied the feasibility of a measurement of the $J/\psi(\rightarrow \mu^+\mu^-)$ differential cross-section as a function of the transverse momentum \cite{CMSjpsi}. The main background comes from events containing two muons mainly from two different decays accidentally having the same invariant mass. The events are selected with a dimuon trigger with a threshold of 3 GeV$/c$ for both muons. Offline, cuts on the invariant mass and the vertex are applied.  

\begin{figure}[h]
\begin{center}
\includegraphics[width=14cm]{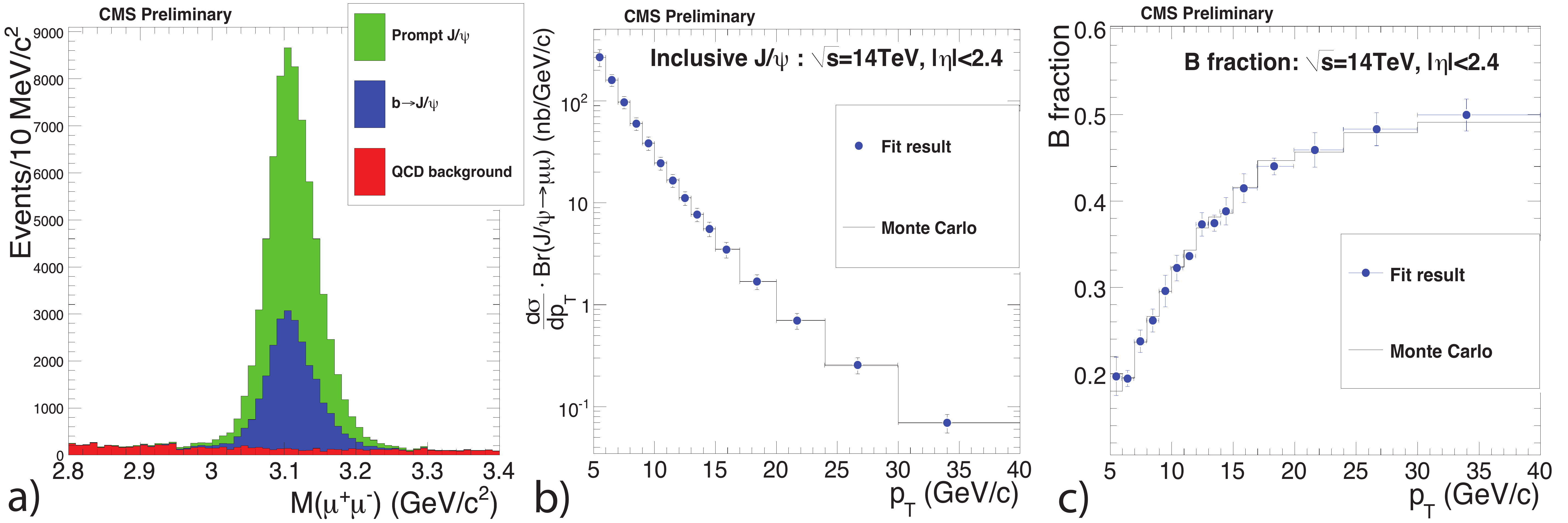}
\end{center}
\caption{a) Invariant mass plot for prompt, non-prompt $J/\psi$'s and background for the full momentum range; b) Inclusive $J/\psi$ cross section; c) Fraction of $J/\psi$'s from $b$-hadron decays. All plots are based on expectations with 3 pb$^{-1}$  of CMS data.}
\label{fig:CMSjpsi}
\end{figure}

The cross section (Figure~\ref{fig:CMSjpsi}b) for each bin in transverse momentum is extracted by fitting the mass spectrum (Figure~\ref{fig:CMSjpsi}a) with a signal and background hypothesis. For 1 pb$^{-1}$ a yield of about 25000 $J/\psi$'s and a mass resolution of approximately 30 MeV$/c^2$ is expected. The fraction of $J/\psi$'s from $b$-hadron decays (Figure~\ref{fig:CMSjpsi}c) is determined with an unbinned likelihood fit on the decay length distribution  for each transverse momentum bin.

\subsection{Quarkonia polarisation measurement}
ATLAS proposes a method to measure the polarisation of the $J/\psi$ and $\Upsilon$ states \cite{csc}. This can be achieved by measuring the angular distribution of the muons from the $J/\psi$ ($\Upsilon$) decay. The angle $\Theta^\star$ is defined as the angle between the $\mu^+$ and the $J/\psi$ ($\Upsilon$) boost direction in the $J/\psi$ ($\Upsilon$) rest frame. The angular distribution is connected to the polarisation parameter $\alpha$ via $	\frac{\text{d} N}{\text{d}\cos \Theta} \propto \left( 1 + \alpha \cos^2 \Theta ^\star \right)$~\cite{csc}. $\alpha$ is equal to +1 for transversely polarised production, -1 for longitudinally polarised production and 0 for unpolarised production. 
The  $J/\psi$ ($\Upsilon$) are reconstructed by using a dimuon trigger with 4 and 6 GeV thresholds for the two muons. Offline, invariant mass and vertex cuts are applied. For the $J/\psi$ the single muon trigger with a threshold of 10 GeV$/c$ is used in addition since the angular acceptance depends highly on the trigger. For the $\Upsilon$ this is not possible due to larger backgrounds. An uncertainty between 0.02 and 0.06 for the $J/\psi$ and about 0.2 for the $\Upsilon$ is reached in a momentum range of $12-21$ GeV$/c$. 


\section{$b$-production studies}
\label{sec:bprod}
Due to the large cross-section for $b$-quark production at the LHC it is of high importance to understand the production processes. There are three production processes: flavour creation, flavour excitation and gluon splitting which have large uncertainties. Furthermore $b$-quark production is the main background to many other analyses, such as Higgs or SUSY searches.

\subsection{Inclusive $b$-production}
CMS proposes to measure the $b$-hadron spectrum by selecting events with a single muon trigger on Level-1 and a muon + $b$-jet trigger in the High-Level-Trigger. A search for the highest transverse momentum $b$-jet is performed offline which requires to have associated muon. The distribution of the relative momentum of the muon with respect to the $b$-jet is used to distinguish between $b$, $c$ and lighter quark jets. The expected uncertainty on the measurement is less than 20\% for a transverse momentum up to 1 TeV$/c$\quad\cite{tdr2}.

\subsection{$B^+ \rightarrow J/\psi K^+$ production}
The decay $B^+ \rightarrow J/\psi \left(\rightarrow \mu^+\mu^-\right) K^+$ has a very clear topology. It is also a reference channel for rare $b$-decays and allows detector studies due to its well known properties. To identify this decay ATLAS uses a single muon trigger with a threshold of 6 GeV$/c$ in transverse momentum. In the offline selection a second muon with a transverse momentum of at least 3 GeV$/c$ is required. To reconstruct the $J/\psi$ cuts on the common vertex and on the invariant mass are applied. An additional track displaced from the primary vertex is required for the $K^+$. Further cuts on the common vertex to the three particles are applied. The number of signal events is determined from a fit on the reconstructed $B^+$ mass. The expected uncertainty for a measurement with 10 pb$^{-1}$ in 5 bins of transverse momentum is expected to be between 15 and 20\%~\cite{csc}.

\subsection{$b\bar{b}$-correlations}

\begin{figure}[h]
\begin{center}
a) \includegraphics[width=6.5cm]{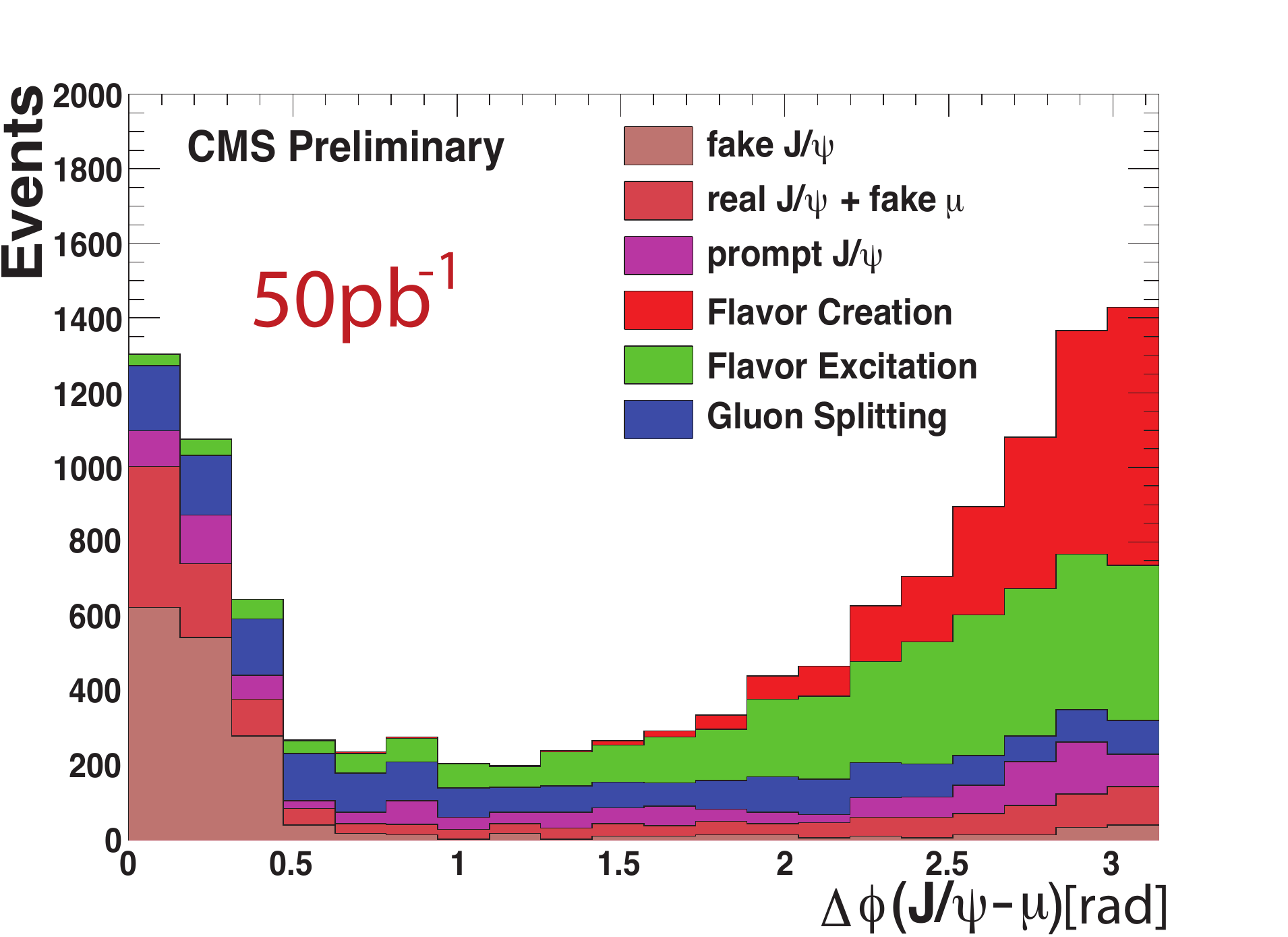}
b) \includegraphics[width=6.5cm]{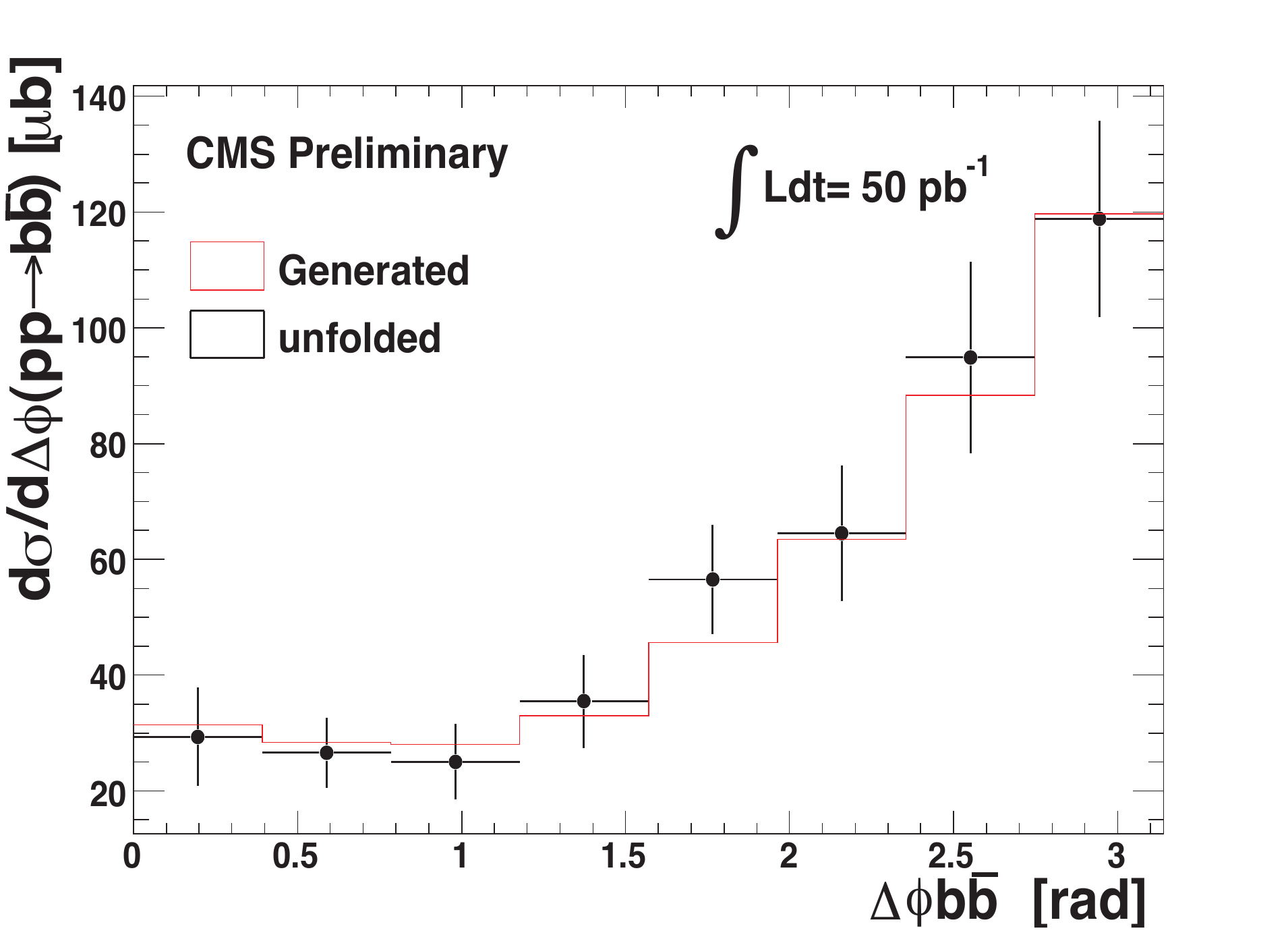}
\end{center}7

\caption{a) Distribution of the angle $\Delta\phi$ between the reconstructed $J/\psi$ and the reconstructed muon. b) Unfolded distribution of the angle between the $b$ and $\bar{b}$. Events were generated with a center-of-mass energy of 10 TeV and for an integrated luminosity of 50 pb$^{-1}$}.
\label{fig:bbcorr}
\end{figure}

Another approach taken by CMS to determine the fraction of the different production mechanisms is to measure the angular correlation between the two $b$-quarks~\cite{CMSbcorr} which depends on the production model. This study was done for a center-of-mass energy of 10 TeV. 
The events are selected with a dimuon trigger with a threshold in transverse momentum of 3 GeV$/c$. The first $b$-hadron is required to decay into a $J/\psi$ whereas for the second only a muon from an arbitrary b-hadron decay is reconstructed. The signal events are extracted with an unbinned likelihood fit on the invariant $J/\psi$ mass, the transverse $J/\psi$ decay length and the distance of closest approach  of the third muon to the beamline. The angular distribution between the $J/\psi$ and the muon (Figure \ref{fig:bbcorr}a) is unfolded  to the angular distribution between the $b$ and $\bar{b}$ (Figure \ref{fig:bbcorr}b) using simulation.

\section{$b$-decay studies}
\label{sec:bdecay}

\subsection{$B_d \rightarrow J/\psi K^*$ and $B_s \rightarrow J/\psi\phi$ decays}

 Both decays, $B_d \rightarrow J/\psi (\rightarrow \mu^+ \mu^-)K^*(\rightarrow K^+ \pi^-)$ and $B_s \rightarrow J/\psi  (\rightarrow \mu^+ \mu^-) \phi(\rightarrow K^+ K^-)$ are very promising for the startup of LHC due to their high rate. They will be an important tool to test the detector calibration and trigger systems. Furthermore the $B_s$-decay opens interesting physics issues giving the possibility to improve the CDF and D0 measurement on the width difference $\Delta\Gamma_s$ between the light and heavy mass eigenstates. 
 
\begin{figure}[h]
\begin{center}
a)\includegraphics[width=7.5cm]{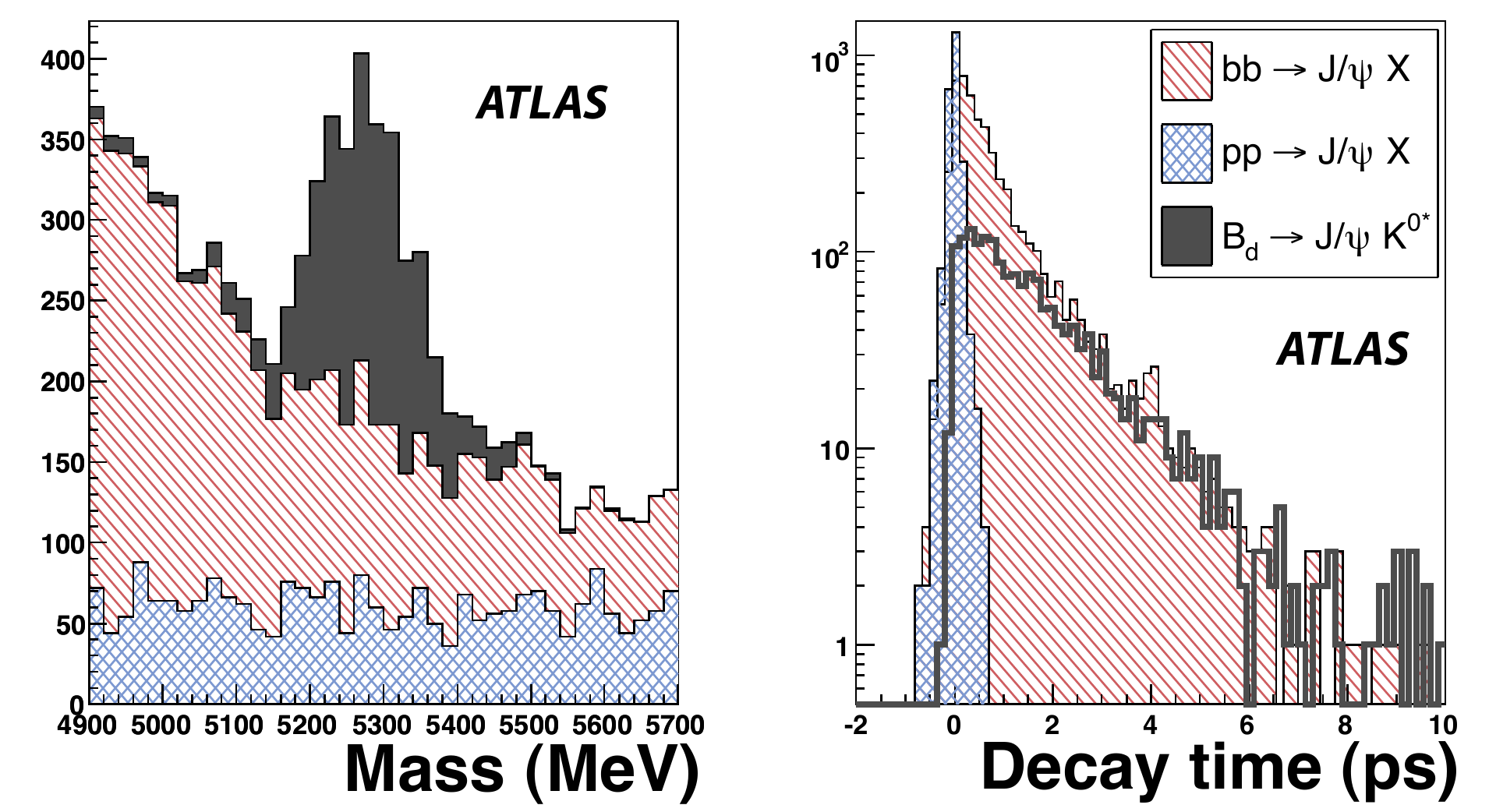}
b)\includegraphics[width=7.5cm]{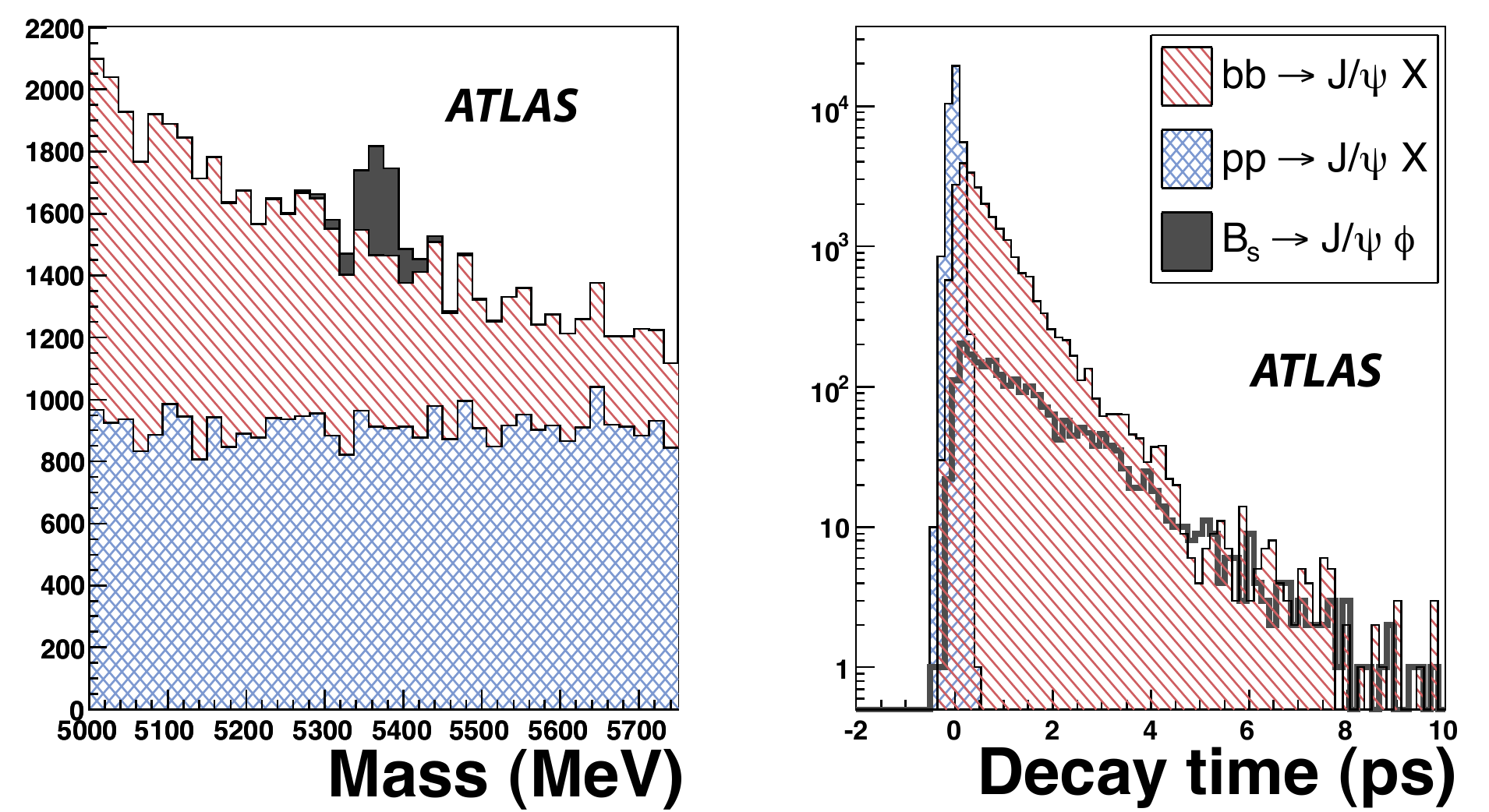}
\end{center}
\caption{a) Invariant mass distribution and decay time spectrum for the $B_d \rightarrow J/\psi K^\star$ decay for an integrated luminosity of 10 pb$^{-1}$; b) Same distributions for the $B_s \rightarrow J/\psi \phi$ decay for an integrated luminosity of 150 pb$^{-1}$.}
\label{fig:Bds}
\end{figure}

 ATLAS developed a strategy for the measurements of $B_d$ ($B_s$) decays with a simple reconstruction algorithms as independent as possible of the reconstruction software~\cite{csc}. The events are selected with a dimuon trigger with transverse momentum thresholds of 4 and 6 GeV$/c$. Then the $J/\psi$ is reconstructed from two muons and the $K^*$ ($\phi$)  from two tracks assuming a kaon and a pion (two kaons). Cuts on vertex and transverse momentum are applied and a simultaneous fit on the invariant mass and decay time is then performed. The invariant mass and decay time is shown in Figure~\ref{fig:Bds} for the $B_d$- and for the $B_s$-decay.
 
CMS proposes a measurement of $\Delta\Gamma_s$ using a tighter selection of the events~\cite{tdr2}. A dedicated trigger with full event reconstruction is used and cuts on the secondary vertex are applied. A kinematic vertex fit is applied offline and an angular analysis is performed to extract the width difference. Assuming a width difference of 20\% an uncertainty of 4\% is expected for 1.3~fb$^{-1}$.

\subsection{Rare decay studies: $B_s \rightarrow \mu\mu$}

The decay $B_s \rightarrow \mu^+\mu^-$ is forbidden at tree level in the standard model resulting in a very low predicted branching ratio of $(3.42 \pm 0.54) \cdot 10^{-9}$. New particles can contribute to the lowest order loop diagrams thereby increasing the branching ratio by orders of magnitude. The events are selected with the already mentioned dimuon triggers, applying offline cuts on muon separation, isolation, decay length and invariant mass. For an integrated luminosity of 10~fb$^{-1}$ both, CMS and ATLAS expect 6 SM signal events and 14 background events~\cite{csc,tdr2}. The 90\% confidence upper limit for $B_s \rightarrow \mu^+\mu^-$ is $1.4 \cdot 10^{-8}$.



\end{document}